\documentclass[epj,twocolumn]{webofc}
\usepackage[varg]{txfonts}   
\usepackage{epsfig}
\usepackage{color}
\newcommand{\ms}[1]{\color{red} #1}
\newcommand{\note}[1]{{\tiny\ms (note)}}
\woctitle{Flavour changing and conserving processes}

\graphicspath{ {figs/} }

\begin{document}
\title{Higher order hadronic and leptonic contributions to the muon $g-2$}
%
%

\author{Alexander Kurz\inst{1,3}\thanks{\email{alexander.kurz2@kit.edu}} \and
        Tao~Liu\inst{2}\thanks{\email{ltao@ualberta.ca}} \and
        Peter~Marquard\inst{3}\thanks{\email{peter.marquard@desy.de}} \and
        Alexander V.~Smirnov\inst{4}\thanks{\email{asmirnov80@gmail.com}} \and
\\
        Vladimir A.~Smirnov\inst{5}\thanks{\email{smirnov@theory.sinp.msu.ru}} \and
        Matthias
        Steinhauser\inst{1}\thanks{\email{matthias.steinhauser@kit.edu}}
}

\institute{Institut f{\"u}r Theoretische Teilchenphysik, Karlsruhe Institute
  of Technology (KIT), 76128 Karlsruhe, Germany
\and
  Department of Physics, University of Alberta, Edmonton AB T6G 2J1, Canada
\and
  Deutsches Elektronen Synchrotron DESY, Platanenallee 6, 15738 Zeuthen, Germany           
\and
  Scientific Research Computing Center, Moscow State University, 119991,
  Moscow, Russia
\and
  Skobeltsyn Institute of Nuclear Physics of Moscow State University, 119991,
  Moscow, Russia
          }

\abstract{%
  In this contribution we discuss next-to-next-to-leading order hadronic
  and four-loop QED contributions to the anomalous magnetic moment of the muon.
}

\maketitle



\section{Introduction}

The anomalous magnetic moment of the muon, $a_\mu$, is among the most precise
measured quantities in particle physics. It is measured to a precision of
0.54~parts per million which matches the precision of the Standard Model
theory prediction~\cite{Bennett:2006fi,Roberts:2010cj}. However, since many
years one observes a discrepancy of about three to four standard deviations
which survives persistent all improvements. This concerns both the
experimental data and theoretical calculations entering the prediction.

Currently a new experiment is built at FERMILAB with
the aim to increase the accuracy of the measured value by about a factor
four~\cite{Carey:2009zzb,Herzog_fccp15}.  In the upcoming years also
improvements on the theory side can be expected. On the one hand this is
connected to improved measurements of $R(s)$ at low energies (see, e.g.,
Refs.~\cite{Aubert:2009ad,Ambrosino:2010bv,Eidelman_fccp15}). On the other
hand it can be expected that within the next few years results from lattice
simulations become available both for the hadronic vacuum polarization and
hadronic light-by-light contributions (see, e.g.,
Refs.~\cite{DellaMorte:2011aa,Burger:2013jya,Blum:2014oka,Blum:2015gfa,Lehner_fccp15}).

The by far dominant numerical contribution to $a_\mu$ originates from QED
corrections which are known to five-loop order~\cite{Aoyama:2012wk}. Note,
however, that the four- and the five-loop corrections have only been computed
by a single group.\footnote{Partial four-loop corrections have been obtained
  in
  Refs.~\cite{Laporta:1993ds,Baikov:1995ui,Aguilar:2008qj,Lee:2013sx,Baikov:2012rr,Baikov:2013ula}.}
For this reason we have recently started to systematically check the four-loop
results of~\cite{Aoyama:2012wk}. In Ref.~\cite{Lee:2013sx} analytic results
for the gauge-invariant subsets with two or three closed electron loops have
been obtained neglecting power corrections of the form $m_e/m_\mu$.  All
contributions involving a $\tau$ lepton have been computed in
Ref.~\cite{Kurz:2013exa}. After including three (analytic) expansion terms in
$m_\mu^2/m_\tau^2$ a better precision has been obtained than in the numerical
approach of Ref.~\cite{Aoyama:2012wk}.  The numerically most important QED contributions
at four-loop level arise from light-by-light-type diagrams (i.e. the external
photon does not couple to the external muon line) containing a closed electron
loop. This well-defined subset has been considered in Ref.~\cite{Kurz:2015bia}
where an asymptotic expansion for $m_e \ll m_\mu$ has been performed to
compute four expansion terms.

We adopt the notation from Ref.~\cite{Aoyama:2012wk} and 
parametrize the anomalous magnetic moment in the form
\begin{eqnarray}
  a_\mu &=& \sum_{n=1}^\infty a_\mu^{(2n)} \left( \frac{\alpha}{\pi} \right)^n
  \,,
  \label{eq::amu}
\end{eqnarray}
where the four-loop contribution can be written as
\begin{eqnarray}
  a_\mu^{(8)} &=& A_1^{(8)} + A_2^{(8)}(m_\mu / m_e) + A_2^{(8)}(m_\mu / m_\tau) 
  \nonumber\\&&\mbox{}
  + A_3^{(8)}(m_\mu / m_e, m_\mu / m_\tau)
  \,.
  \label{eq::Amu}
\end{eqnarray}
$A_1^{(8)}$ contains only contributions from photons and muons,
$A_2^{(8)}(m_\mu / m_e)$ and $A_2^{(8)}(m_\mu / m_\tau)$ involve closed
electron or tau loops, and each Feynman diagram which contributes to
$A_3^{(8)}(m_\mu / m_e, m_\mu / m_\tau)$ contains all three lepton flavours
simultaneously.  In Sections~\ref{sec::electron} and~\ref{sec::tau} we
describe the calculation of the light-by-light type QED contribution to
$A_2^{(8)}(m_\mu / m_e)$ (see also~\cite{Kurz:2015bia}) and the computation of
$A_2^{(8)}(m_\mu / m_\tau)$ (see also~\cite{Kurz:2013exa}), respectively.
Afterwards we summarize in Section~\ref{sec::hadr} the computation of the
next-to-next-to-leading order (NNLO) hadronic vacuum polarization contribution
published in Ref.~\cite{Kurz:2014wya}.  A brief summary and an outlook is
given in Section~\ref{sec::summary}.


\section{\label{sec::electron}Four-loop electron contribution}

\begin{figure*}[t]
  \begin{center}
    \mbox{}\hfill
    \includegraphics[scale=0.8]{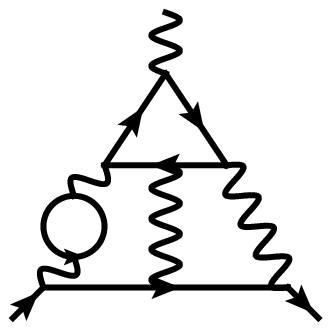}
    \hfill
    \includegraphics[scale=0.8]{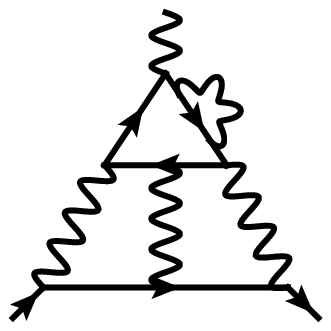}
    \hfill
    \includegraphics[scale=0.8]{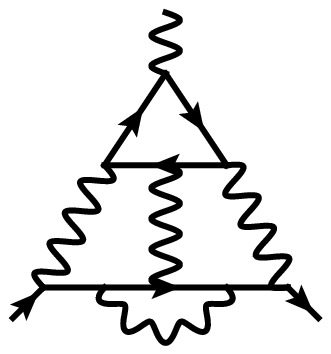}
    \hfill\mbox{}
    \\[1em]
    \mbox{} \hspace*{8em}
    IV(a) \hfill IV(b) \hfill IV(c)
    \hspace*{8em} \mbox{}
  \end{center}
  \caption{\label{fig::FDs}Sample light-by-light-type Feynman diagrams
    contributing to $a_\mu$. The external solid line represents the muon and at least
    one of the internal solid loops denotes electrons. In the case of IV(a) the
    second fermion loop can either be an electron or muon loop. Wavy lines
    represent photons.}
\end{figure*}

The numerically most important contribution to $a_\mu^{(8)}$ originates from
diagrams involving a closed electron loop (denoted by $A_2^{(8)}(m_\mu / m_e)$
in Eq.~(\ref{eq::Amu}). This contribution contains a gauge invariant subset
where the external photon does not couple to the external muon line but to a
closed fermion loop, the so-called leptonic light-by-light-type diagrams. Due
to Furry's theorem such diagrams do not contribute at two but only start at
three loops where four photons can be attached to the closed fermion
loop. Here we discuss the four-loop result which can be sub-divided into three
gauge invariant and finite contributions which we denote by IV(a), IV(d) and
IV(c). Sample Feynman diagrams are shown in Fig.~\ref{fig::FDs}. Case IV(a) can be
further subdivided according to the flavour of the leptons in the closed
fermion loops.  The contribution with two electron loops is denoted by IV(a0),
the one with one muon and one electron loop and the coupling of the external
photon to the electron by IV(a1), and the remaining one with one muon and one
electron loop by IV(a2). We do not consider the case with two muon loops since
this contribution is part of $A_1^{(8)}$.

The light-by-light-type diagrams are numerically dominant and provide about
95\% of the four-loop electron loop contribution. The main reason for this are
$\log(m_e/m_\mu)$ terms which are even present in the limit $m_e\to0$. In fact,
IV(a0) even has quadratic logarithms which makes this part the most
important one.

Our calculation is based on an asymptotic
expansion~\cite{Beneke:1997zp,Smirnov:2013} for $m_e\ll m_\mu$ which is
implemented with the help of {\tt asy}~\cite{Pak:2010pt,Jantzen:2012mw} and
in-house {\tt Mathematica} programs. Similar to the hard mass procedure applied in
Section~\ref{sec::tau} we obtain a factorization of the original two-scale
integrals into products of one-scale integrals. The latter are either vacuum
or on-shell integrals or integrals containing eikonal propagators of the form
$1/(p\cdot q)$ (see Ref.~\cite{Kurz:2015bia} for more details).  For each
integral class we perform a reduction to master integrals and obtain analytic
results expressed as a linear combination
of about 150 so-called master integrals. About 50\% of them we know
analytically or to high numerical precision. The remaining ones are computed
with the help of the package {\tt FIESTA}~\cite{Smirnov:2013eza} which is the
source of the numerical uncertainty in our final result.  We would like to
stress that in our approach a systematic improvement is possible if it is
required to improve the accuracy.

\begin{table}[t]
  \begin{center}
  \begin{tabular}{l|r|r}
    $A_2^{(8)}\left(\frac{m_\mu}{m_e}\right)$ & 
    \multicolumn{1}{c|}{\cite{Kurz:2015bia}} 
    & \multicolumn{1}{c}{\cite{Kinoshita:2004wi,Aoyama:2012wk}} \\
    \hline
    IV(a0) & $116.76 \pm 0.02$ & $116.759183  \pm 0.000292$ \\
    IV(a1) & $2.69 \pm 0.14$   & $2.697443 \pm 0.000142$  \\
    IV(a2) & $4.33 \pm 0.17$  & $4.328885 \pm 0.000293$  \\
    IV(a) & $123.78\pm 0.22$ & $123.78551\hphantom{2} \pm 0.00044\hphantom{2}$  \\
    IV(b) & $-0.38 \pm 0.08$ & $-0.4170\hphantom{22}   \pm 0.0037\hphantom{22}$ \\
    IV(c) & $2.94 \pm 0.30$  & $2.9072\hphantom{22}    \pm 0.0044\hphantom{22}$ \\
  \end{tabular}
  \caption{\label{tab::a8}Summary of the final results for the individual
    four-loop light-by-light-type contributions and their comparison with 
    results presented in Refs.~\cite{Kinoshita:2004wi,Aoyama:2012wk}.}
  \end{center}
\end{table}

For all five cases we compute terms up to order $(m_e/m_\mu)^3$ (i.e. four
expansion terms) and check that the cubic corrections only provide a
negligible contribution. Our final results can be found in Tab.~\ref{tab::a8}
where we compare to the findings of
Refs.~\cite{Kinoshita:2004wi,Aoyama:2012wk}.  Note that results for IV(a0)
have also been obtained in Refs.~\cite{Calmet:1975tw,Chlouber:1977dr}, though
with significantly larger uncertainty.  In all cases good agreement is found
with~\cite{Kinoshita:2004wi,Aoyama:2012wk}. Although our numerical
uncertainty, which amounts to approximately $0.4 \times (\alpha/\pi)^4 \approx
1.2 \times 10^{-11}$, is larger, the final result is nevertheless sufficiently
accurate as can be seen by the comparison to the difference between the
experimental result and theory prediction which is given by
\begin{eqnarray}
  a_\mu({\rm exp}) - a_\mu({\rm SM}) &\approx& 249(87) \times 10^{-11}
  \,.
  \label{eq::amu_diff}
\end{eqnarray}
This result is taken from Ref.~\cite{Aoyama:2012wk}. Note that the uncertainty
in Eq.~(\ref{eq::amu_diff}) receives approximately the same amount from
experiment and theory. Even after a projected reduction of the uncertainty
by a factor four both in $a_\mu({\rm exp})$ and $a_\mu({\rm SM})$ our
numerical precision is a factor ten below the uncertainty of the difference.


\section{\label{sec::tau}Four-loop tau lepton contribution}

In this section we discuss the gauge invariant and finite subset of Feynman
diagrams involving a closed heavy tau lepton loop. In the limit of infinitely
heavy $m_\tau$ this contribution has to vanish. Thus $A_2^{(8)}(m_\mu /
m_\tau)$ has a parametric dependence $m_\mu^2/m_\tau^2$ which is of order $10^{-3}$.
Note, that $\alpha/\pi \approx 2\cdot 10^{-3}$ and thus one can expect that
the four-loop tau lepton contribution is of the same order as 
the universal five-loop result~\cite{Aoyama:2012wk}.

We compute this contribution by applying an asymptotic expansion in the limit
$m_\tau^2 \gg m_\mu^2$. This is realized with the help of the program {\tt
  exp}~\cite{Harlander:1997zb,Seidensticker:1999bb} which is written in {\tt
  C++}.  As a result the two-scale four-loop integrals factorize into
one-scale vacuum ($m_\tau$) and on-shell ($m_\mu$) integrals.  Both integral
classes are well studied in the literature (for references
see~\cite{Kurz:2013exa}). This concerns both the reduction to master integrals
and the analytic evaluation of the latter.

\begin{figure*}[t]
  \begin{center}
    \begin{tabular}{c}
      \leavevmode
      \epsfxsize=0.85\textwidth
      \epsffile[80 490 550 670]{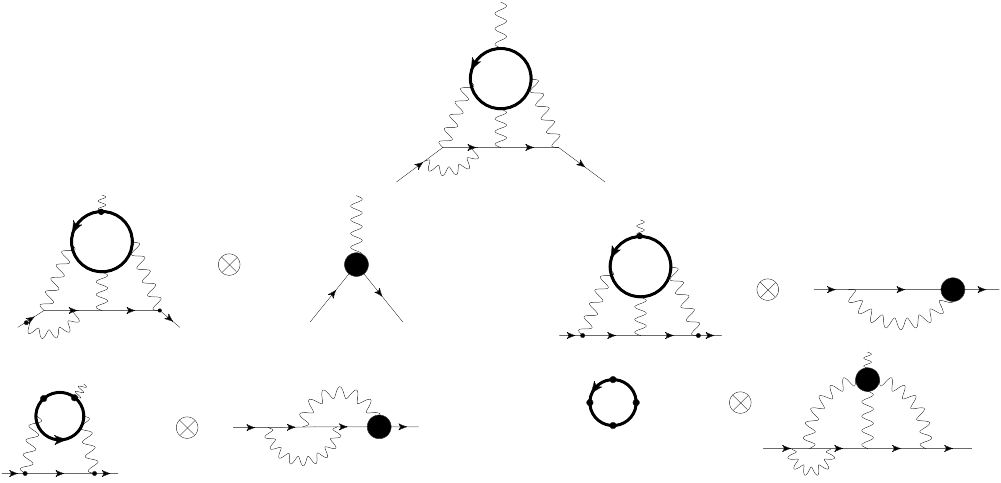}
    \end{tabular}
  \end{center}
  \caption[]{\label{fig::ae} Graphical example for the application of the
    asymptotic expansion at four loops. Thick solid, thin
    solid and wavy lines represent taus, muons and photons,
    respectively. Only four representative sub-diagrams are
    shown; altogether there are eight contributions for the diagram
    in the first row.} 
\end{figure*}

In the first line of Fig.~\ref{fig::ae} a sample Feynman diagram is shown
where the thick solid lines represent the tau leptons. Rows two and three of
Fig.~\ref{fig::ae} show the result of the asymptotic expansion where the
graphs left of the symbol $\otimes$ have to be expanded in all small
quantities, i.e., the external momenta and the muon mass. Thus, the only mass
scale of the remaining vacuum integral is the tau lepton mass.  The result of
the Taylor expansion is inserted into the effective vertex (thick blob)
present in the diagram to the right of $\otimes$. Afterwards the remaining
loop integrations, which are of on-shell type, are performed.

As a final result we obtain an expansion in $m_\mu^2/m_\tau^2$ with analytic
coefficients containing $\log(m_\mu^2/m_\tau^2)$ terms. 
Note that with the help of this method a better accuracy has been obtained than
with the numerical approach of Ref.~\cite{Aoyama:2012wk}.
Inserting numerical values for the lepton masses leads to
\begin{eqnarray}
  A^{(8)}_{2,\mu}(m_\mu/m_\tau) &=&
  0.0421670 + 0.0003257 
  \nonumber\\&&\mbox{}
  + 0.0000015 + \ldots
  \,,
  \label{eq::A8}
\end{eqnarray}
where the ellipsis indicates terms of order $(m_\mu^2/m_\tau^2)^4$ which are
expected to contribute at order $10^{-8}$ to $A^{(8)}_{2,\mu}(m_\mu/m_\tau)$.

$a_\mu$ receives contribution from $\tau$ lepton loops starting at two-loop
order. Their numerical impact is given by
\begin{eqnarray}
  10^{11} \times a_\mu\Big|_{\tau \rm loops}
  &=& 42.13 + 0.45 + 0.12
  \,,
  \label{eq::amu_tau}
\end{eqnarray}
where the numbers on the right-hand side correspond to the two, three and four
loops.  It is interesting to note that the three-loop term is only less than a
factor four larger than the four-loop counterpart. Furthermore, it is worth
comparing the numbers in Eq.~(\ref{eq::amu_tau}) to the universal
contributions contained in $A_{1}$ which read~\cite{Aoyama:2012wk}
\begin{eqnarray}
  10^{11} \times a_\mu\Big|_{\rm univ.} \!\!\!&=&\!\!\! 
  116\,140\,973.21 - 177\,230.51 
  \nonumber\\&&\mbox{}
  + 1\,480.42 - 5.56 + 0.06
  \,,
\end{eqnarray}
where the individual terms on the right-hand side represent the results from
one to five loops.  Note that the four-loop tau lepton term is twice bigger
than the five-loop photonic contribution.


\section{\label{sec::hadr}NNLO hadronic contribution}

\begin{figure*}[t]
  \begin{center}
  \begin{tabular}{cccc}
    \includegraphics[scale=0.65]{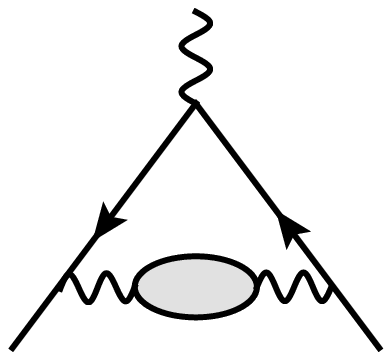} &
    \includegraphics[scale=0.65]{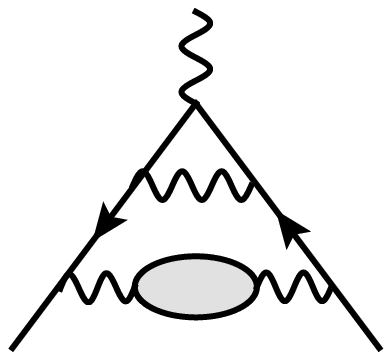} &
    \includegraphics[scale=0.65]{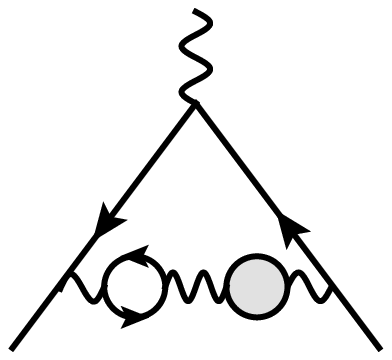} &
    \includegraphics[scale=0.65]{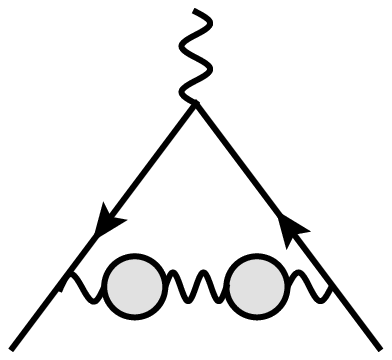}
    \\ (a) LO & (b) $2a$ & (c) $2b$ & (d) $2c$ \\
    \includegraphics[scale=0.65]{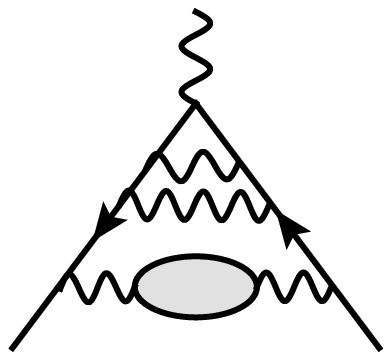} &
    \includegraphics[scale=0.65]{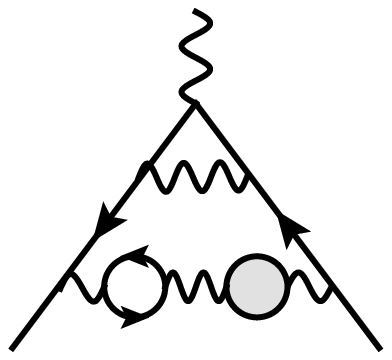} &
    \includegraphics[scale=0.65]{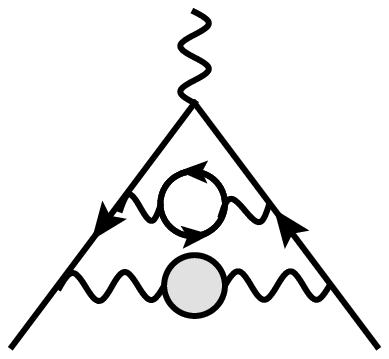} &
    \includegraphics[scale=0.65]{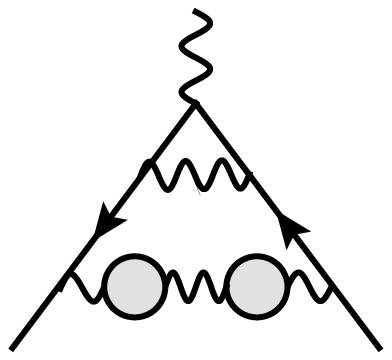}
    \\ (e) $3a$ & (f) $3b$ & (g) $3b$ & (h) $3c$ \\
    \includegraphics[scale=0.65]{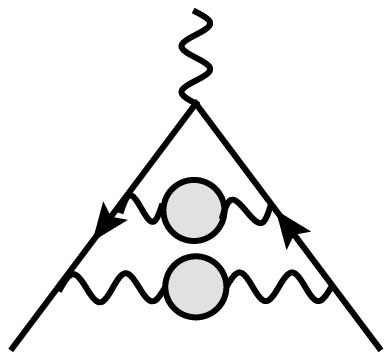} &
    \includegraphics[scale=0.65]{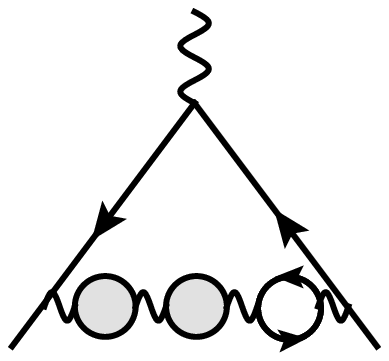} &
    \includegraphics[scale=0.65]{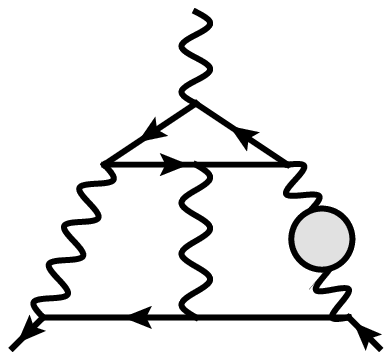} &
    \includegraphics[scale=0.65]{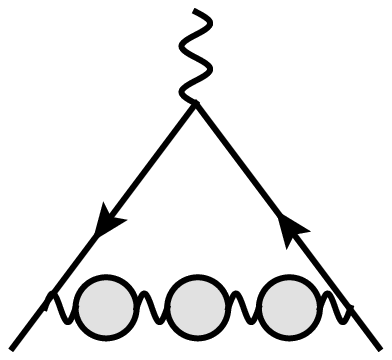}
    \\ (i) $3c$ & (j) $3c$ & (k) $3b$,lbl & (l) $3d$ \\
  \end{tabular}
  \caption{\label{fig::FD_nnlo}Sample LO, NLO and NNLO Feynman diagrams
    contributing to $a_\mu^{\rm had}$. The external fermions are muons and the
    fermions in the closed loops represent electrons.}
  \end{center}
\end{figure*}

The LO hadronic contribution to the anomalous magnetic moment of the muon
is obtained from diagram (a) in Fig.~\ref{fig::FD_nnlo}.
One parametrizes the hadronic contribution (represented by the blob)
by the polarization function $\Pi(q^2)$ which appears as a factor in the
integrand of the one-loop diagram. In a next step one exploits analyticity of
$\Pi(q^2)$ and uses a dispersion integral to introduce its imaginary part,
\begin{eqnarray}
  R(s) &=& \frac{ \sigma(e^+e^-\to\mbox{hadrons}) }{ \sigma_{pt} }
  \,,
\end{eqnarray}
with $\sigma_{pt} = 4\pi\alpha^2/(3s)$. Note that 
$\sigma(e^+e^-\to\mbox{hadrons})$ does note include initial state
radiative or vacuum polarization corrections.
At that point the loop integration and the dispersion integral are interchanged
and one obtains
\begin{eqnarray}
  a_\mu^{(1)} &=& \frac{1}{3} \left(\frac{\alpha}{\pi}\right)^2
  \int_{m_\pi^2}^\infty {\rm d} s \frac{R(s)}{s} K^{(1)}(s)
  \,,
  \label{eq::aLO}
\end{eqnarray}
A convenient integral representation for the kernel function $K^{(1)}(s)$, 
which is the result of the loop integration, is given by
\begin{eqnarray}
  K^{(1)}(s) &=& 
  \int_0^1 {\rm d} x \frac{x^2(1-x)}{x^2 + (1-x) \frac{s}{m_\mu^2}}
  \,.
  \label{eq::K1}
\end{eqnarray}
At one-loop order it is possible to obtain analytic results (see
Refs.~\cite{BroRaf68,Eidelman:1995ny}). Nevertheless, it is promising to
consider $K^{(1)}(s)$ in the limit $m_\mu^2 \ll s$ which is justified since
the lower integration limit in Eq.~(\ref{eq::aLO}) is $m_\pi^2$ which is
bigger than $m_\mu^2$. The expansion of $K^{(1)}(s)$ is easily obtained by
remembering that it originates from the vertex diagram similar to
Fig.~\ref{fig::FD_nnlo}(a) where the hadronic blob (including the external
photon lines) is replaced by a massive photon with mass $\sqrt{s}$. The
expansion $m_\mu^2 \ll s$ is easily implemented with the help of the program
{\tt exp}~\cite{Harlander:1997zb,Seidensticker:1999bb} which implements the
rules of asymptotic expansions involving a large internal mass (see, e.g.,
Ref.~\cite{Smirnov:2013}). As a result the original two-scale integral is
represented as a sum of one-scale integrals which are simple to compute.
Using this approach several expansion terms in $m_\mu^2/s$ can be computed.
One observes that an excellent approximation for $a_\mu^{(1)}$ is obtained by
including terms up to order $(m_\mu^2/s)^5$.

The approach described in detail for the one-loop diagram can also be applied
at two and three loops where exact calculations of the kernel functions are
either very difficult or even impossible. In Ref.~\cite{Kurz:2014wya} four
expansion terms have been computed which provides an approximation at the per
mil level.

A slight complication arises for the contributions involving more than one
hadronic insertion, see Figs.~\ref{fig::FD_nnlo}(d,h,i,j,l). In case they are
present in the same photon line formulas similar to Eq.~(\ref{eq::K1}) can be
derived with two- and three-dimensional integrations. Diagrams of type $(3c)$
in Fig.~\ref{fig::FD_nnlo} are more involved. Here, we apply a multiple
asymptotic expansion in the limits $s\gg s^\prime \gg m_\mu^2$, $s \approx
s^\prime\gg m_\mu^2$ and $s^\prime\gg s\gg m_\mu^2$ ($s$ and $s^\prime$ are
the integration variables) and construct an interpolating function by
combining the results from the individual limits.

The LO result for the hadronic vacuum polarization contribution to $a_\mu$ can
be found in
Refs.~\cite{Davier:2010nc,Hagiwara:2011af,Jegerlehner:2011ti,Benayoun:2012wc,Jegerlehner:2015stw}
and NLO analyses have been performed in
Refs.~\cite{Krause:1996rf,Greynat:2012ww,Hagiwara:2003da,Hagiwara:2011af}.
Our NLO results for the three contributions read
\begin{eqnarray}
  a_\mu^{(2a)} &=& -20.90 \times 10^{-10}\,, \nonumber\\
  a_\mu^{(2b)} &=&  10.68 \times 10^{-10}\,, \nonumber\\
  a_\mu^{(2c)} &=&   0.35 \times 10^{-10}\,,
\end{eqnarray}
which leads to
\begin{eqnarray}
  a_\mu^{\rm had,NLO} &=& -9.87 \pm 0.09 \times 10^{-10}\,,
\end{eqnarray}
in a good agreement with Refs.~\cite{Hagiwara:2003da,Hagiwara:2011af}.  Note
that in our analyses no correlated uncertainties are taken into account.  Such
a rough treatment should not be done at LO but is certainly acceptable at NNLO.

For the individual NNLO contributions we obtain the results
\begin{eqnarray}
  a_\mu^{(3a)} &=&         0.80   \times 10^{-10} \,,\nonumber\\
  a_\mu^{(3b)} &=&        -0.41   \times 10^{-10} \,,\nonumber\\
  a_\mu^{(3b,\rm lbl)} &=& 0.91   \times 10^{-10} \,,\nonumber\\
  a_\mu^{(3c)} &=&        -0.06   \times 10^{-10} \,,\nonumber\\
  a_\mu^{(3d)} &=&         0.0005 \times 10^{-10} \,,
\end{eqnarray}
which leads to
\begin{eqnarray}
  a_\mu^{\rm had,NNLO} &=& 1.24 \pm 0.01 \times 10^{-10}\,.
  \label{eq::amuNNLO}
\end{eqnarray}
It is interesting to note that similar patterns are observed at two and three
loops: multiple hadronic insertions are small and the contributions of type
(b) involving closed electron two-point functions reduce the contributions of
type (a) by about 50\%.  However, at three-loop order there is a new type of
diagram where the external photon couples to a closed electron loop
($a_\mu^{(3b,\rm lbl)}$) which provides the largest individual
contribution. This is in analogy to the three-loop QED corrections where the
light-by-light type diagrams dominate the remaining contributions.
In fact, due to $a_\mu^{(3b,\rm lbl)}$ the NNLO hadronic vacuum polarization
contribution has a non-negligible impact. It has the same order of magnitude
as the current uncertainty of the leading order hadronic contribution and
should thus be included in future analyses.

An important contribution to $a_\mu$ is provided by the so-called hadronic
light-by-light diagrams where the external photon is connected to the hadronic
blob. The NLO part of this contribution is of the same perturbative order as
the corrections in Eq.~(\ref{eq::amuNNLO}). A first-principle calculation of
this part is currently not available, however, in~\cite{Colangelo:2014qya} it
has been estimated to 
$a_\mu^{\rm lbl-had,NLO}=0.3 \pm 0.2 \times 10^{-10}$.

We want to mention that there is a further hadronic contribution where
four internal photons couple to the hadronic blob and the external photon
couples to the muon line (``internal hadronic light-by-light'').
This contribution, which is formally of the same perturbative order
as $a_\mu^{\rm had,NNLO}$, is currently unknown.


\section{\label{sec::summary}Summary and conclusions}

For more than a decade
the measured and predicted results for the anomalous magnetic moment of the
muon show a discrepancy of three to four standard deviations.
This circumstance has triggered many publications which try to
interpret the deviation with the help of beyond-SM theories.
However, before drawing definite conclusions it is necessary to
cross check the experimental result by performing an independent high-precision
determination of $a_\mu$. Furthermore, all ingredients of the theory
prediction should be computed by at least two groups independently.

In this contribution we describe the calculation of two classes of four-loop QED
contributions to $a_\mu$, which up to date only have been computed by one
group: the contribution involving tau leptons and the one involving
light-by-light-type closed electron loops. Good agreement with the results in
the literature is found.  To complete the cross check of the four-loop result
the non-light-by-light electron contribution, the diagrams involving
simultaneously electrons and taus, and the pure-muon contribution have to be
computed.  From the technical point of view the missing diagram classes have
the same complexity as those described in Sections~\ref{sec::electron}
and~\ref{sec::tau}.

As a further topic we have discussed in Section~\ref{sec::hadr} the
calculation of the NNLO hadronic vacuum polarization contribution.


\section*{Acknowledgments}

M.S. would like to thank the organizers of ``Flavour changing and conserving
processes 2015'' for the pleasant atmosphere during the conference.  We thank
the High Performance Computing Center Stuttgart (HLRS) and the Supercomputing
Center of Lomonosov Moscow State University~\cite{LMSU} for providing
computing time used for the numerical computations with {\tt FIESTA}.  P.M. was
supported in part by the EU Network HIGGSTOOLS PITN-GA-2012-316704.  This work
was supported by the DFG through the SFB/TR~9 ``Computational Particle
Physics''.  The work of V.S. was supported by the Alexander von Humboldt
Foundation (Humboldt Forschungspreis).


%
%

\end{document}